\documentclass{article}

\usepackage{arxiv}

\usepackage[utf8]{inputenc} 
\usepackage[T1]{fontenc}    
\usepackage{hyperref}       
\usepackage{url}            
\usepackage{booktabs}       
\usepackage{amsfonts}       
\usepackage{nicefrac}       
\usepackage{microtype}      
\usepackage{lipsum}		
\usepackage{graphicx}

\usepackage{amsmath}
\usepackage{chemformula}

\title{Deep learning for thermal plasma simulation: solving 1-D arc model as an example}

\author{ Linlin Zhong\thanks{This paper is currently under consideration by a journal. The author's homepage:\href{https://mathboylinlin.com}{https://mathboylinlin.com}}, ~Qi Gu, and Bingyu Wu \\
	School of Electrical Engineering\\
	Southeast University\\
	No 2 Sipailou, Nanjing, Jiangsu Province 210096, P. R. China\\
	\texttt{mathboylinlin@gmail.com}, \texttt{linlin@seu.edu.cn}\\
}

\date{February 12, 2020}

\renewcommand{\headeright}{}
\renewcommand{\undertitle}{A Preprint}

\begin{document}
\maketitle

\begin{abstract}
Numerical modelling is an essential approach to understanding the behavior of thermal plasmas in various industrial applications. We propose a deep learning method for solving the partial differential equations in thermal plasma models. In this method a deep feed-forward neural network is constructed to surrogate the solution of the model. A loss function is designed to measure the discrepancy between the neural network and the equations describing thermal plasmas. A good neural network is obtained by minimizing this loss function. We demonstrate the power of this deep learning method by solving a 1-D arc decaying model which is consist of three cases: stationary arc, transient arc without considering radial velocity, and transient arc with radial velocity respectively. The results show that the deep neural networks have excellent ability to express the differential equations describing thermal plasmas. This could bring us a new and prospective numerical tool for thermal plasma modelling.
\end{abstract}


\section{Introduction}
Thermal plasmas are partially or strongly ionized gases which are usually produced by electric arcs, RF discharges, microwave discharges, and laser induced plasmas \cite{gleizes2005thermal}. There are a lot of industrial applications relating to thermal plasmas, such as arc welding, arc lighting, circuit interruption, plasma cutting, plasma spraying, mineral processing, nanoparticle production, and waste treatment \cite{murphy2016modeling}. In order to well understand thermal plasmas, numerical modelling is commonly used as an essential tool to predict the properties of thermal plasmas in the above applications. For example, the computational fluid dynamics (CFD) modelling can help those who are studying and developing gas-blast circuit breakers \cite{gonzalez2012turbulence, yang2013low, zhang2014modelling} and arc welding processes \cite{murphy2017desktop, park2017coupled, chen2020model} to understand the flow features of interrupting arc and optimize import welding parameters respectively. The basic equations describing the behavior of thermal plasmas are conservation or continuity equations and Maxwell’s equations \cite{murphy2001thermal}. The former includes the equations of mass continuity, momentum conservation, and energy conservation, which are similar to those used in conventional fluid dynamics. The latter is used to describe the flow of charge. Furthermore, in some cases such as high-velocity plasma jet and supersonic nozzle in high-voltage circuit breakers, extra equations are needed to describe the turbulence in thermal plasmas \cite{murphy2001thermal, gleizes2014perpectives}. All these equations are partial differential equations (PDEs) which are solved numerically with defined initial and boundary conditions by the finite element method (FEM) or the finite volume method (FVM). In the practical applications, FVM is much more widely used than FEM for thermal plasma simulation. In order to describe thermal plasmas e.g. the compressive plasma flow with shock waves as accurately as possible, the sophisticated meshing techniques should be applied to computational domain. Even so, the FEM or FVM solvers do not always give us reasonable results because of convergence problem.

\paragraph{}
With the development of artificial intelligence and machine learning in the last decades, we nowadays have a new powerful tool for solving differential equations via deep learning in form of deep neural networks (DNNs) \cite{han2018solving, sirignano2018dgm, tripathy2018deep, raissi2019physics}. DNNs can be used as a black box method to approximate a physical system, which is called physics-informed neural networks by some researchers \cite{raissi2019physics, pang2019fpinns}. The philosophy of learning differential equations through neural networks is not a new idea \cite{lagaris1998artifical, lagaris2000neural} but recent experience has shown that deep networks with many layers seem to do a surprisingly good job in modeling complicated datasets \cite{han2018solving}. The key step of deep learning-based method for solving differential equations is to constrain a neural network which can minimizes the residual of the corresponding differential equations. Solving differential equations is therefore reduced to solving optimization problems. Although we still cannot well explain the effectiveness of neural networks, the universal approximation theorems tell us that hidden-layer neural networks can approximate a wide class of functions on compact subsets \cite{han2018solving}. Compared with the traditional mesh-based methods e.g. FEM and FVM, the deep learning-based method is a mesh-free approach which could break the curse of dimensionality \cite{pang2019fpinns}. 
 
\paragraph{} 
The rest of the paper is organized as follows. In Section \ref{sec:sec2}, we describe in detail a general deep learning-based algorithm for solving partial differential equations in thermal plasma models. In Section \ref{sec:sec3}, we demonstrate the power of this algorithm by solving a 1-D arc decaying model with three cases: stationary arc, transient arc without considering radial velocity, and transient arc with radial velocity respectively. Lastly a short summary is presented in Section \ref{sec:sec4}.

\section{Methodology}
\label{sec:sec2}

In this work we propose a deep learning method for thermal plasma modelling by taking advantages of recent developments in artificial intelligence, such as automatic differentiation and acceleration of graphics processing unit (GPU). Since the equations used to describe thermal plasmas are partial differential equations defined on spatio-temporal domain $\Omega$, we consider the following general PDE parameterized by $\lambda$ for the solution $u(t, \mathbf{x})$ with $\mathbf{x} = (x, y, z)$.
\begin{equation}
\label{equ:equ1}
	f(t,x,y,z,\frac{\partial u}{\partial t}, \frac{\partial u}{\partial x}, \frac{\partial u}{\partial y}, \frac{\partial u}{\partial z}, \frac{\partial^2 u}{\partial x^2}, \frac{\partial^2 u}{\partial y^2}, \frac{\partial^2 u}{\partial z^2}, \frac{\partial^2 u}{\partial xy}, \frac{\partial^2 u}{\partial xz}, \frac{\partial^2 u}{\partial yz}, \mathbf{\lambda}) = 0
\end{equation}

With boundary and initial conditions defined on $\partial \Omega$

\begin{equation}
\label{equ:equ2}
	\mathcal{B}(u,t,\mathbf{x}) = 0, \ \mathcal{I}(u,\mathbf{x}) = 0
\end{equation}

Where the parameters $\mathbf{\lambda}$ represent thermal plasma properties, e.g. mass density, specific heat, electrical conductivity, thermal conductivity, and viscosity. They are always temperature and pressure dependent. The meaning of solution u depends on the type of equations. For example, $u$ stands for velocity and pressure in momentum conservation equations and temperature in energy conservation equation.

\begin{figure}
	\centering
	\includegraphics[width=9.5cm]{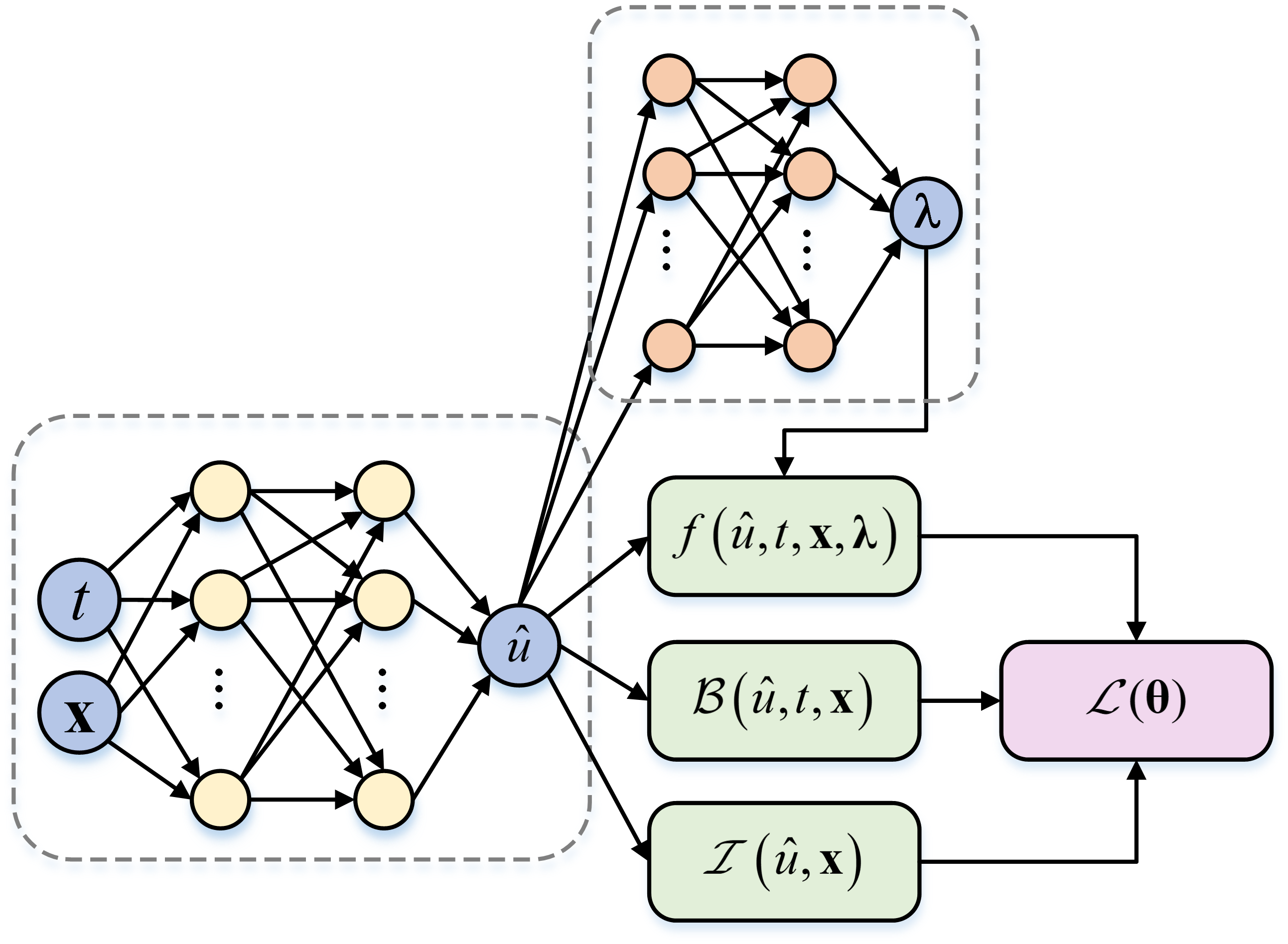}
	\caption{Diagram of a deep neural network for thermal plasma modelling. A large multi-layer feed-forward neural network is used to surrogate the solution of differential equations describing thermal plasmas. A smaller neural network is used to express the plasma properties in the differential equations.}
	\label{fig:fig1}
\end{figure}

\paragraph{} 
As illustrated in Figure \ref{fig:fig1}, in order to solve the equation (\ref{equ:equ1}), we construct a feed-forward neural network $\hat{u}(t,\mathbf{x},\mathbf{\theta})$ with parameters $\mathbf{\theta}$ as the surrogate of $u(t, \mathbf{x})$. This network takes $t$ and $\mathbf{x}$ as input, and outputs the predicted solution of $u$. The parameters $\mathbf{\theta}$ are the set of weights in the neural network, which are initialized randomly or using some specific techniques e.g. Xavier initialization method \cite{glorot2010understanding} before the training of the neural network. Definitely, the neural network $\hat{u}(t,\mathbf{x},\mathbf{\theta})$ in the initial stage does not satisfy the differential equation (\ref{equ:equ1}) and boundary and initial conditions (\ref{equ:equ2}). In order to restrict $\hat{u}(t,\mathbf{x},\mathbf{\theta})$ to satisfy these constraints, we construct a loss function $\mathcal{L}(\mathbf{\theta})$ as follows to measure the discrepancy between $\hat{u}(t,\mathbf{x},\mathbf{\theta})$ and the constraints.

\begin{equation}
\label{equ:equ3}
	\mathcal{L}(\mathbf{\theta}) = \mathcal{L}_f(\mathbf{\theta}) + \mathcal{L}_\mathcal{B}(\mathbf{\theta}) + \mathcal{L}_\mathcal{I}(\mathbf{\theta})
\end{equation}

\begin{equation}
\label{equ:equ4}
	\mathcal{L}_f(\mathbf{\theta}) = \frac{1}{N_f} \sum_{i=0}^{N_f} \mathcal{F}(f(t_i, x_i, y_i, z_i, \frac{\partial \hat{u}_i}{\partial t_i}, \frac{\partial \hat{u}_i}{\partial x_i}, \frac{\partial \hat{u}_i}{\partial y_i}, \frac{\partial \hat{u}_i}{\partial z_i}, \frac{\partial^2 \hat{u}_i}{\partial x_i^2}, \frac{\partial^2 \hat{u}_i}{\partial y_i^2}, \frac{\partial^2 \hat{u}_i}{\partial z_i^2}, \frac{\partial^2 \hat{u}_i}{\partial x_iy_i}, \frac{\partial^2 \hat{u}_i}{\partial x_iz_i}, \frac{\partial^2 \hat{u}_i}{\partial y_iz_i}, \mathbf{\lambda}))
\end{equation}

\begin{equation}
\label{equ:equ5}
	\mathcal{L}_\mathcal{B}(\mathbf{\theta}) = \frac{1}{N_\mathcal{B}} \sum_{i=0}^{N_\mathcal{B}} \mathcal{F}(\mathcal{B}(\hat{u}_i, t_i, \mathbf{x}_i))
\end{equation}

\begin{equation}
\label{equ:equ6}
	\mathcal{L}_\mathcal{I}(\mathbf{\theta}) = \frac{1}{N_\mathcal{I}} \sum_{i=0}^{N_\mathcal{I}} \mathcal{F}(\mathcal{I}(\hat{u}_i, \mathbf{x}_i))
\end{equation}

Where $N_f$,  $N_\mathcal{B}$, and $N_\mathcal{I}$ are the number of the scattered points we select in the corresponding computational domain $\Omega$ and the boundary and initial domain $\partial \Omega$. These points are called training data which can be selected randomly or uniformly. All the partial derivatives in (\ref{equ:equ4}) - (\ref{equ:equ6}) are calculated by automatic differentiation during the training \cite{raissi2019physics}. $\mathcal{F}(\cdot)$ is the function for measuring residuals. All the previous works used $L^1$ or $L^2$ norm as $\mathcal{F}(\cdot)$ \cite{sirignano2018dgm, tripathy2018deep, raissi2019physics}. In this work we combine the advantages of $L^1$ and $L^2$ norm, and use Huber loss function as follows instead based on our practical experience.

\begin{equation}
\label{equ:equ7}
	\mathcal{F}(f(x)) = \begin{cases}
	\frac{1}{2} (f(x))^2 & \mbox{for } \left|f(x)\right| \leq 0 \\
	\delta \left|f(x)\right| - \frac{1}{2}\delta^2 & \mbox{otherwise}
	\end{cases}
\end{equation}

Where $\delta$ is a controlling parameter which is set as 1 in this work.

\paragraph{} 
For the parameter $\mathbf{\lambda}$ representing thermal plasma properties, it is difficult to take it into account directly as in the previous works \cite{sirignano2018dgm, tripathy2018deep, raissi2019physics} because the thermodynamic, transport, and radiation properties of a thermal plasma are usually given in a tabular form not in the explicit functional form. Moreover, these plasma properties are strongly dependent of temperature and pressure i.e. $\hat{u}$ in the neural network. As a result, we propose to use several pre-trained small neural networks to express the relationship between $\hat{u}$ and $\mathbf{\lambda}$. This is similar to the commonly used regression techniques.

\paragraph{} 
Lastly, we need to search for good weights $\mathbf{\theta}$ of the neural network by minimizing the loss function $\mathcal{L}(\mathbf{\theta})$, which is called “training” in deep learning. The well-known and widely used algorithm Adam \cite{kingma2014adam} is used to accomplish this non-convex optimization problem in this work.

\section{Case study of 1-D arc modely}
\label{sec:sec3}

Next, we demonstrate the power of this deep learning-based method by solving 1-D arc decaying model as an example which is consist of three cases: stationary arc, transient arc without considering radial velocity, and transient arc with radial velocity respectively. All the works were implemented using the framework of PyTorch which is an open source and popular machine learning library developed by Facebook's AI Research lab (FAIR) \cite{paszke2019pytorch}.

\subsection{Case 1: stationary arc}
\label{sec:sec3.1}

In the first case, we construct a 6-hidden-layer deep neural network with 50 neurons per layer to solve the Elenbaas-Heller equation which describes the radial temperature distribution of a steady arc in a cylindrical column \cite{shaw2006regular, liao2016approximate}.

\begin{equation}
\label{equ:equ8}
	\frac{1}{r} \frac{\partial}{\partial r}(r\kappa \frac{\partial T}{\partial r}) + \sigma \frac{I^2}{g^2} - E_{rad} = 0
\end{equation}

\begin{equation}
\label{equ:equ9}
	g = \int_0^R 2\pi r \sigma \, dr
\end{equation}

Where $r$ is the radial distance, $T$ the arc temperature, $\sigma$ the electrical conductivity, $\kappa$ the thermal conductivity, $I$ the current, $g$ the arc conductance, $E_{rad}$ the energy loss due to radiation, and $R$ the radius of the arc. The radiation energy loss is described by the net emission model \cite{lowke1973decay} in this work. For the boundary conditions, the zero gradient of temperature i.e. $\partial T/\partial r = 0$ and the constant temperature (2000 K in this work) are used at the symmetry ($r = 0$) and the wall ($r = R$) respectively.

\begin{figure}
	\centering
	\includegraphics[width=9.5cm]{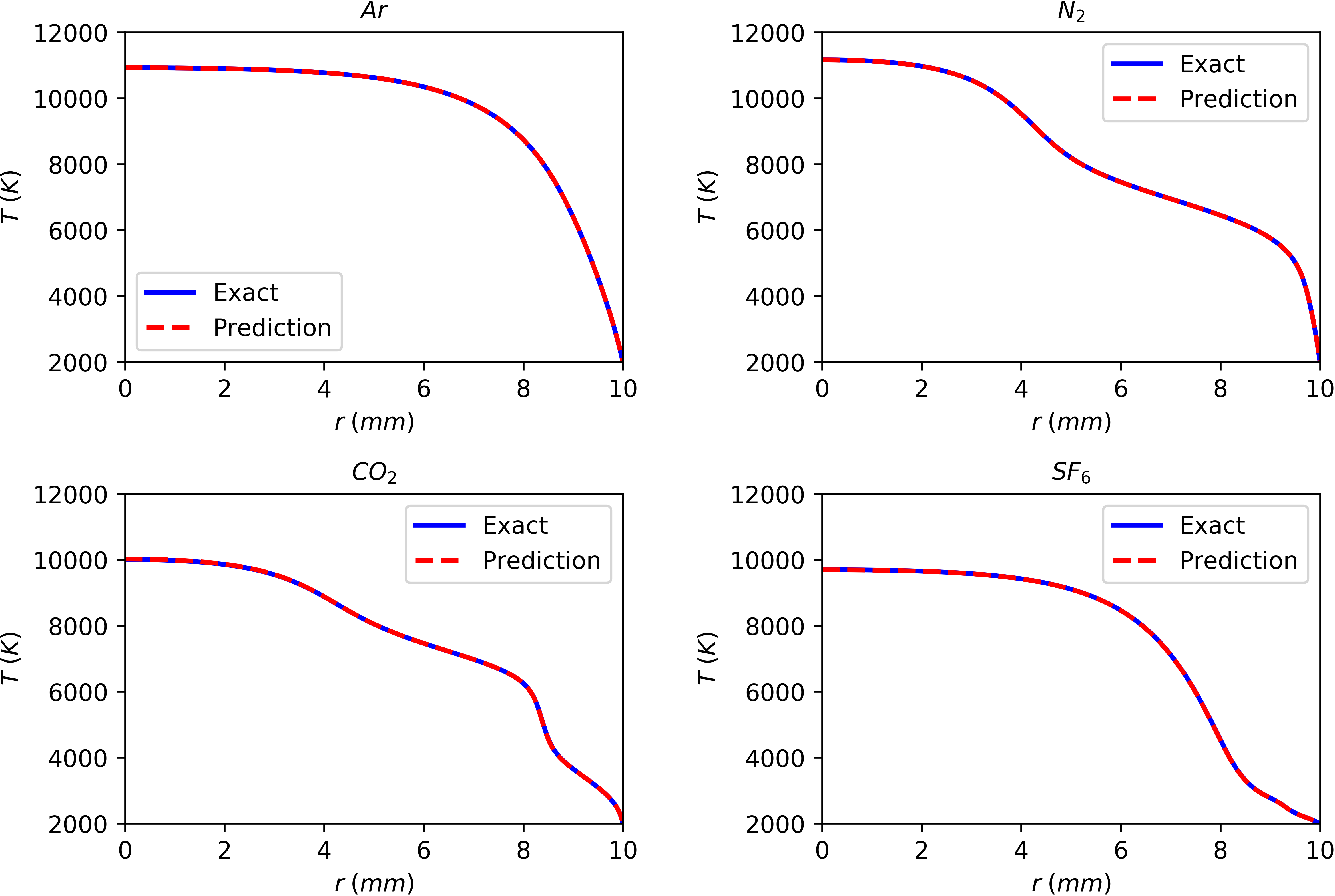}
	\caption{Prediction of arc temperature distribution of \ch{Ar}, \ch{N_2}, \ch{CO_2}, and \ch{SF_6} arc plasmas at ambient pressure ($I$ = 200A, $R$ = 10mm) via deep learning neural networks with comparison of the exact results by a high-order BVP solver.}
	\label{fig:fig2}
\end{figure}

\paragraph{}
Since the electrical conductivity, thermal conductivity, and the net emission coefficients in the radiation model are all dependent of arc temperature, we use 6-hidden-layer deep neural networks with 20 neurons per layer to fit the functional relationship between temperature and these plasma properties. All the source data are compiled from our previous works \cite{zhong2019evaluation, zhong2019improved, zhong2019effects}. The integral in (\ref{equ:equ9}) is calculated by the Gauss-Legendre quadrature. The hyperbolic tangent function is used as activation functions in all the neural networks. 500 points are sampled uniformly along the arc radius as training data. The learning rate is set to be $10^{-4}$ during the whole training. The Elenbaas-Heller equation is generally difficult to solve analytically because of the strong nonlinear characteristics of the electrical conductivity \cite{shaw2006regular}. In order to validate the results by the deep learning, we also solve the solution as the exact results by a high-order boundary value problem (BVP) solver \cite{kierzenka2008bvp}. The comparison is shown in Figure \ref{fig:fig2}. As can be seen, the radial temperatures of \ch{Ar}, \ch{N_2}, \ch{CO_2}, and \ch{SF_6} arc plasmas solved by deep neural networks agree very well with the exact results even in the domain near the boundary wall in which the temperature gradients could be very large.

\subsection{Case 2: transient arc without radial velocity}
\label{sec:sec3.2}

In the second case, we add a transient term in the equation which describes a decaying arc with time. This usually exists in the extinguishing process of interrupting arc in gas-blast circuit breakers \cite{zhong2019evaluation, zhong2019improved}.

\begin{equation}
\label{equ:equ10}
	\rho C_p \frac{\partial T}{\partial t} = \frac{1}{r} \frac{\partial}{\partial r} (r \kappa \frac{\partial T}{\partial t}) + \sigma \frac{I^2}{g^2} - E_{rad}
\end{equation}

Where $\rho$ and $C_p$ are the mass density and the specific heat at constant pressure respectively. Both of them are also temperature dependent and are also expressed by 6-hidden-layer deep neural networks with 20 neurons per layer as in the first case. Due to the transient term, an initial condition is needed in addition to the boundary conditions in the first case. We use the steady temperature profile obtained in the first case as the initial values of temperature. The current is set to be zero after that ($t > 0$).

\begin{figure}
	\centering
	\includegraphics[width=12cm]{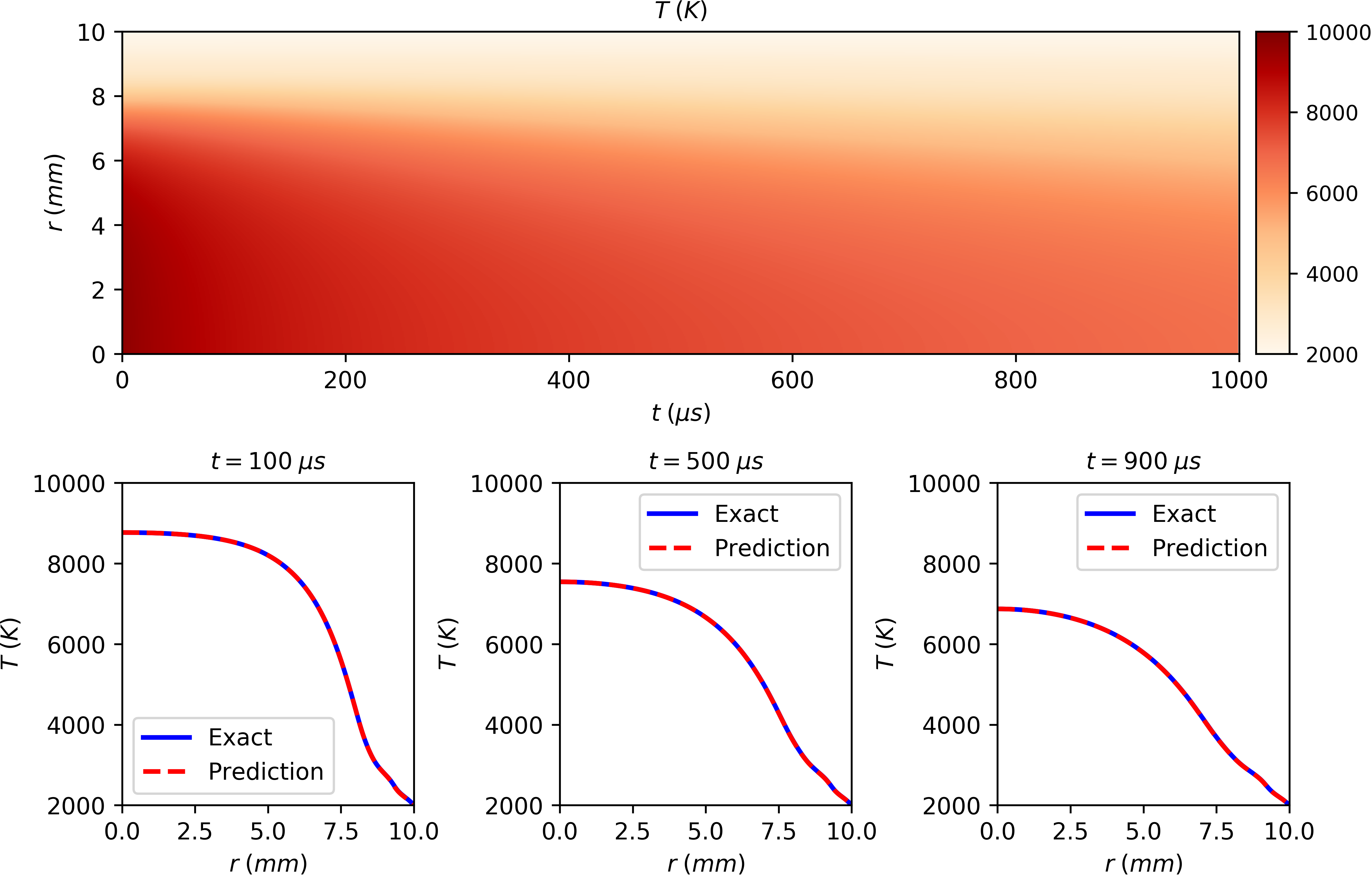}
	\caption{Prediction of the spatial and temporal distribution of arc temperature in transient \ch{SF_6} arc plasma without considering radial velocity at ambient pressure ($R$ = 10mm, $t$: 0 $\sim$ 1000$\mu$s) via deep learning neural networks with comparison of the exact results by an explicit method.}
	\label{fig:fig3}
\end{figure}

\paragraph{}
Also due to the transient term, a larger neural network than that in the first case is needed to surrogate the temperature of transient arc. In this work, 6-hidden-layer deep neural network with 200 neurons per layer is used in solving the equation \ref{equ:equ10}. 200 points along the arc radius and 100 points along the temporal axis are sampled uniformly, which totally results in 20000 training data ($t, \mathbf{x}$) in the computational domain $\Omega$. Also, 200 and 100 points are sampled uniformly at both spatial and temporal boundary $\partial \Omega$.  All the other settings are consistent with the first case. In order to obtain accurate results for comparison, we follow the work by Chervy et al. \cite{chervy1996calculation} and use an explicit method with very short time step of 1 ns. Figure \ref{fig:fig3} presents the spatial and temporal distribution of arc temperature in \ch{SF_6} arc plasma calculated by the deep learning method. The radial temperature distributions at $t$ = 100$\mu $s, 500$\mu$s and 900$\mu$s are compared with the exact results by the explicit method. Obviously, the deep neural network yields very good results both in spatial and temporal domain.

\subsection{Case 3: transient arc with radial velocity}
\label{sec:sec3.3}

In the last case, we take into account the effect of radial velocity $v_r$, which leads to an extra equation of mass continuity and an addition of convection term in the equation of energy conservation.

\begin{equation}
\label{equ:equ11}
	\frac{\partial \rho}{\partial t} + \frac{1}{r} \frac{\partial}{\partial r} (r \rho v_r) = 0
\end{equation} 

\begin{equation}
\label{equ:equ12}
	\rho C_p (\frac{\partial T}{\partial t} + v_r\frac{\partial T}{\partial r}) = \frac{1}{r}\frac{\partial}{\partial r}(r\kappa \frac{\partial T}{\partial r}) + \sigma \frac{I^2}{g^2} - E_{rad}
\end{equation} 

\begin{figure}
	\centering
	\includegraphics[width=12cm]{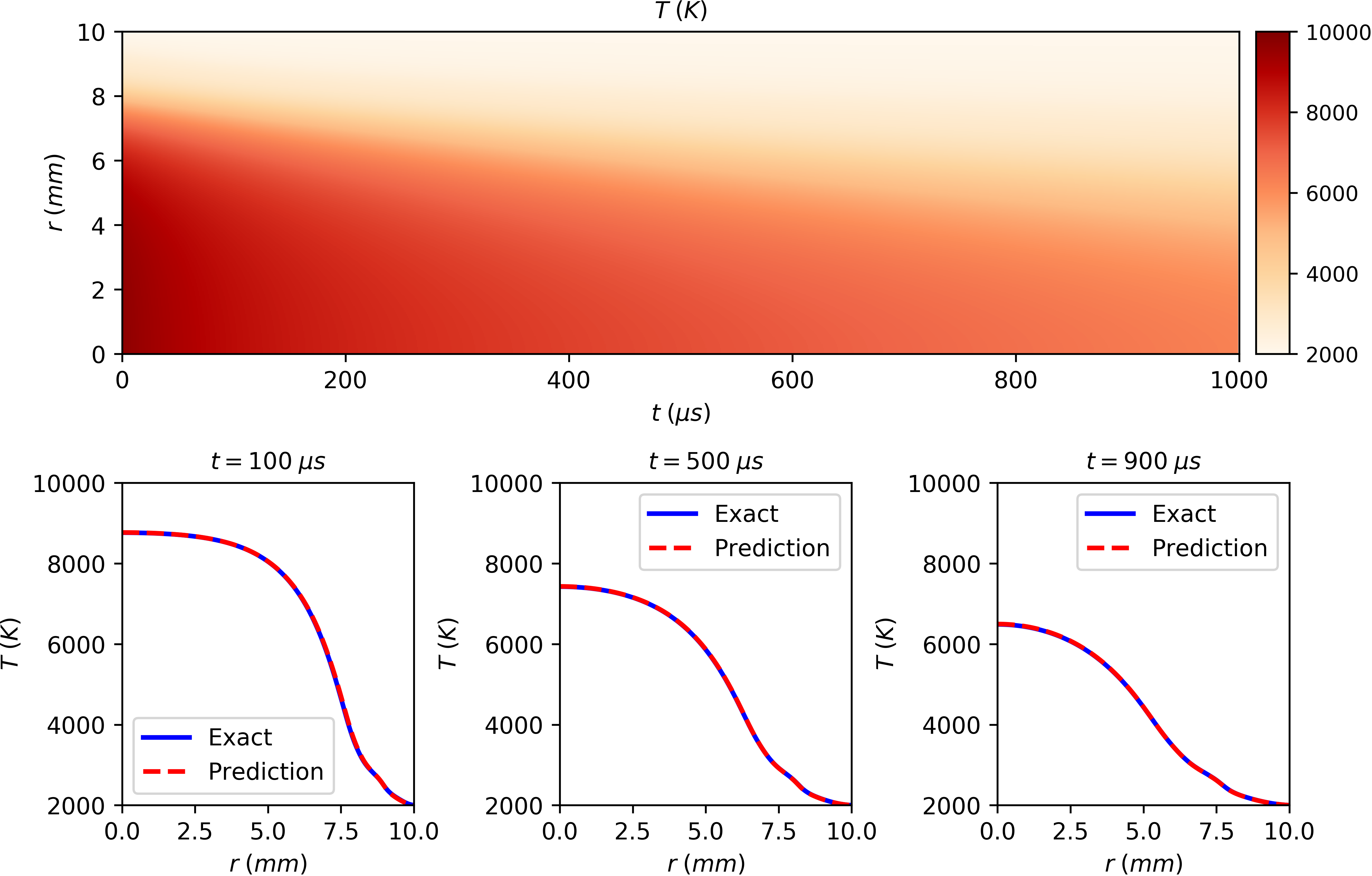}
	\caption{Prediction of the spatial and temporal distribution of arc temperature in transient \ch{SF_6} arc plasma with considering radial velocity at ambient pressure ($R$ = 10mm, $t$: 0 $\sim$ 1000$\mu$s) via deep learning neural networks with comparison of the exact results by an explicit method.}
	\label{fig:fig4}
\end{figure}

\begin{figure}
	\centering
	\includegraphics[width=12cm]{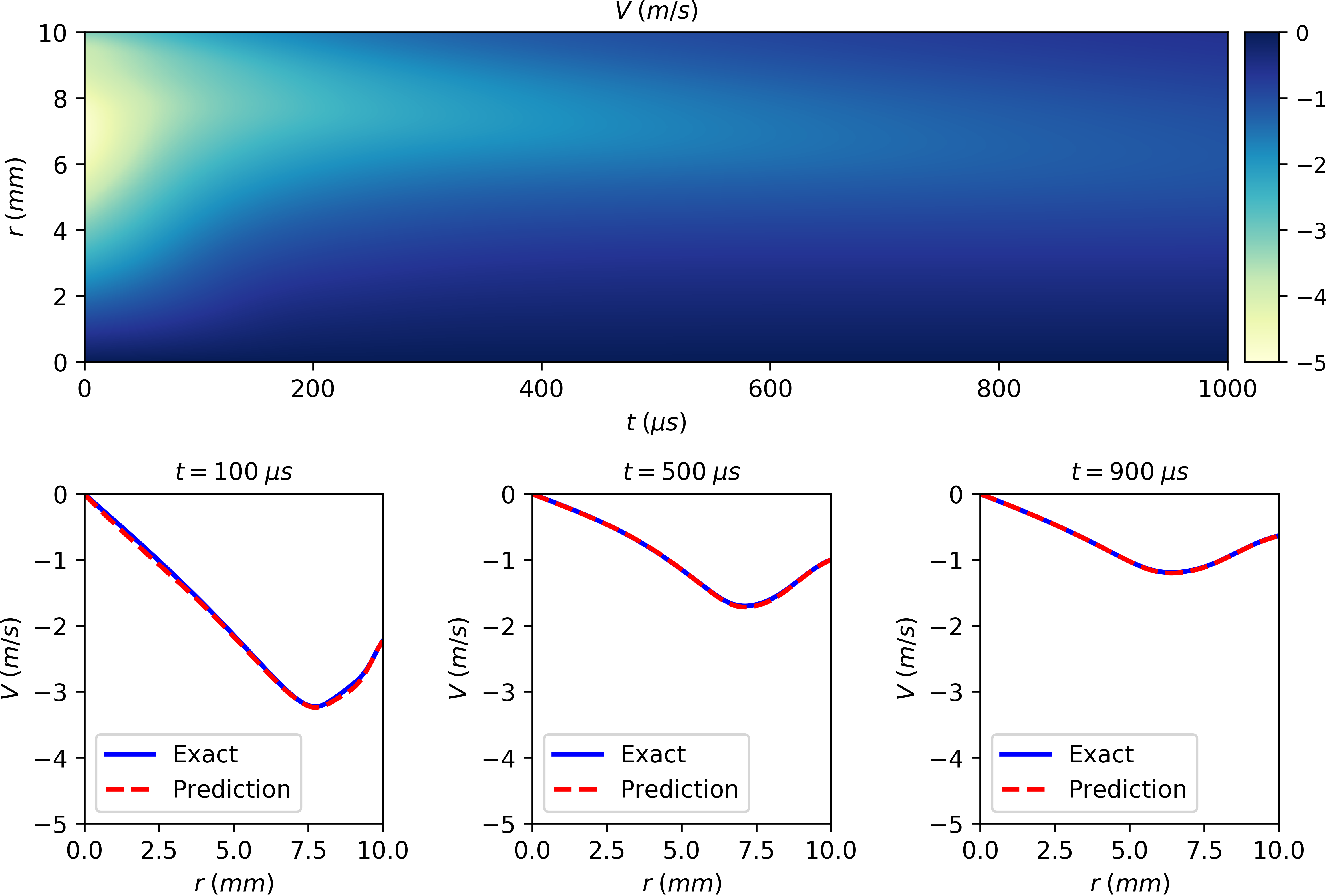}
	\caption{Prediction of the spatial and temporal distribution of radial velocity in transient \ch{SF_6} arc plasma at ambient pressure ($R$ = 10mm, $t$: 0 $\sim$ 1000$\mu$s) via deep learning neural networks with comparison of the exact results by an explicit method.}
	\label{fig:fig5}
\end{figure}

\paragraph{}
This case is more complicated than the second case because the two differential equations must be solved in a coupled way. Therefore, we use a larger neural network than that in the second case, i.e. a 6-hidden-layer deep neural network with 300 neurons per layer. There are two outputs in the last layer of this neural network, corresponding to the temperature $T$ and velocity $v_r$ respectively. The initial velocity ($t = 0$) and the velocity at the symmetry ($r = R$) are both zero. All the other settings are consistent with the second case. The results of temperature and velocity for \ch{SF_6} arc plasma are shown in Figure \ref{fig:fig4} and \ref{fig:fig5} respectively. It can be seen from the comparison between the results by the deep learning method and the explicit method in both figures that the deep neural network has excellent ability to yield good prediction of temperature and velocity for the 1-D arc decaying model.

\section{Conclusion}
\label{sec:sec4}

In summary, we propose a deep learning method for thermal plasma simulation. More specifically, we use deep neural networks to solve the partial differential equations describing thermal plasmas. The power of deep learning in this kind of task is demonstrated preliminarily by solving a 1-D arc decaying model. This could bring plasma community a new and prospective numerical tool for thermal plasma modelling. Following this philosophy, this deep learning-based method could be extended to more complicate thermal plasma modelling in the future work.

\section*{Acknowledgments}
\label{sec:acknowledgments}

This work was supported in part by the National Natural Science Foundation of China (No. 51907023) and the Natural Science Foundation of Jiangsu Province (No. BK20180387).

\bibliographystyle{unsrt}


\end{document}